\documentclass[12pt]{article} 
\newcommand{\be}{\begin{equation}} 
\newcommand{\ee}{ \end{equation}}

\newfont{\msbm}{msbm9 at 11pt}
\newfont{\msbms}{msbm5 at 8pt}

\newfont{\smsbm}{msbm10 at 9pt}
\newfont{\gmsbm}{msbm10 at 18pt}
\newfont{\Gmsbm}{msbm10 at 24pt}

\begin {document}
\begin{center}
 {\bf \Large  On the Generalized Exclusion Statistics}
 \footnote{\it pacs number: 05.30.-d/11.10.-z/11.30.Ly }\\
\vskip 2cm
{\bf  M.Rachidi, E.H.Saidi}\\
UFR-PHE, Facult\'e des Sciences, D\'epartement de Physique, Av. Ibn Battouta B.P. 1014, Rabat, Morocco\\
 and\\
 {\bf J.Zerouaoui} \footnote{\it e-mail: j.zerouaoui@caramail.com }\\
LPTA, Laboratoire de Physique Th\'eorique et Appliqu\'ee, Facult\'e des Sciences, D\'epartement de 
 Physique, B.P. 133, Kenitra, Morocco.\\  
UFR-PHE, Facult\'e des Sciences, D\'epartement de Physique, Av. Ibn Battouta B.P. 1014, Rabat, Morocco.\\
 
   The Abdus Salam International Centre for Theoretical Physics,\\
   Trieste, Italy.\\
\end{center}
\vskip 3cm
\begin{abstract}
\vskip 1cm
We review the principal steps leading to drive the wave function $\psi _{\{k_1,k_2,...,k_N \}}(1,2,...,N)$ 
of a gaz of $N$ identical particle states with exotic statistics. For spins 
$s=1/M$ $mod(1)$, we show that the quasideterminant conjectured in [19], by using 
$2d$ conformal field theoretical methods, is indeed related to the quantum determinant 
of noncommutative geometry. The q-number $[N]!=\prod _{n=1}^N(\sum _{j=0}^{N-1} q^j)$ carrying 
the effect of the generalized Pauli exclusion principle, according to which no more 
than $(M-1)$ identical particles of spin $s=1/M$ $mod(1)$ can live altogether on 
the same quantum state, is rederived in rigourous from the q-antisymmetry. Other 
features are also given.
\end{abstract}
\newpage
\section{Introduction}
Statistics is a quantum mechanical feature of identical particles which governs 
the quantum behavior of a collection of $N$ particles in the large $N$ limite. 
In the $(1+3)$ dimentional real world where experiments are availiable both at 
low and high energy physics, theory and experiments agree on the Pauli classification 
of particles into bosons and fermions respectively of integer and half odd integer 
spins. Outside the real world where experiments are not yet availiable (except for 
very special condensed matter systems [1]), there are only theoretical predictions. 
For higher dimensions, representations group theory predict that like in $(1+3)d$ world, 
the Pauli exclusion principle should usualy hold. In two dimensions however, theory 
predicts that one can go beyond the Pauli classification although one should worry 
about locality if one approach exotic statistics by using $2d$ quantum field theoretical 
methods [2].\\
As far as exotic statistics is concerned and though it was considered in some occasions 
in the past [3], we think that its right development was initiated only in the begining 
of the eighties, especialy after the work of Belavin et al [4] on $2d$ conformal 
field theory (CFT). Since this development, exotic statistics has been attracting 
interest of attension as they are involved in different areas of theoretical physics [5,6,7]. 
Besides its natural generalisation of the Bose and Fermi statistics, one recalls 
for instance the role its plays in conformal minimal models and thier integrable 
deformations [6]. Exotic statistics interpolating between bosons and fermions plays 
also a centrale role in anyon superconductivity, fractional supersymmetry [8,9],
$U_q(sl(2))$ and affine $U_q(s\hat l (2))$ quantum groups representations [7] and 
non commutative geometry [10]. The latter is getting more and more importance in 
superstring theory, especially after the development made by Connes et al [11] in 
the context of the compactification of the matrix models of $M$ theory [12], and too 
recently in type $IIB$ on $AdS_5 \times S^5 /Z_3$ and $4d$ $N=1$ supersymmetric 
$su(N)^3$ gauge theory with bifundamental matters [13]. Applications of exotic 
statistics are then multiple as it is involved in many areas of modern physics. 
One of the these applications which has been considered recently and which be the 
subject of this paper is that consediring the thermodynamic properties of a quantum 
ideal gas of particles obeying exotics statistics. In this contex there are different approches 
dealing with such kind of statistical systems. One of these approaches is the interesting 
one initiated by Haldane [14] and developed by Wu [15], where it was shown that 
the mean occupation number $<n_k>$ of states of definite energy $\epsilon _k$ is given by
\be
<n_k>=\frac{1}{w(e^{\beta (\epsilon _k-\mu)})-\alpha }.
\ee
In this equation, $\beta =1/(kT)$ and $w(\xi)$ is a function obeying the following 
identity
\be
w(\xi)^\alpha [1+w(\xi)]^{1-\alpha }=\xi =e^{\beta (\epsilon _k-\mu)}.
\ee
The parameter $\alpha$ describes the interpolating statistics between bosons 
($\alpha =0$ and $w(\xi)=\xi$) and fermions ($\alpha =1$ and $w(\xi)=\xi -1$). 
An alternative approach dealing with ideal gases of identical particles of exotic 
statistics was also considered by one of us in [16] and developed later in [17,18,19]. 
In this approach, one considers particle states of spin $1/M$ $(mod 1)$; $M\geq 2$ as 
basic objects on the same footing as bosons and fermions associated with $M=\infty$ and 
$M=2$ respectively. This consideration is essentialy motivated by the fact that 
particles of fractional spin $(1/M)$ appears as well definite state in $2d$ fractional 
supersymmetry and in the periodic representations of the quantum group $U_q(s\hat l (2))$ 
with $q$ a root of unity i.e $q=exp(\frac {2i\pi}{M})$ [8,9,20]. One of the remarkable 
results of our way of doing is the statement of the principle governing particles of spin 
$1/M$ $(mod 1)$. Such principle extends the standard Pauli exclusion and tells 
that no more than $(M-1)$ particles of spin $1/M$ $(mod 1)$ can live altogether 
on the same quantum state. In connection with this results, it was shown in [16] 
that the mean occupation number $<n_k>$ describing the quantum behaviour of such exotic 
statistics is given by\\
\be
<n_k>=\frac{1}{e^{-\beta (\epsilon _k-\mu)}-1}-\frac {M}{e^{-\beta M(\epsilon _k-\mu)-1}}.
\ee
Taking $M=\infty$ and $M=2$ in eq.(3), one discovers the usual Bose-Einstein and 
Fermi-Dirac distributions respectively. Eq.(3) was first established in [16] and 
at that time the author of [16] had no idea on the Haldane proposal. The search for the 
distribution (3) was motivated only by studies on perafermions \'a la Zamolodchikov 
and Fateev [21], fractional superstrings \'a la Tye et al [22] and integrable deformation 
of conformal field theory [23]. It should be noted here that in our analysis we start 
from the idea that particle states of fractional spin exist as individual states and 
thier effects should be a priori visible if $2d$ experiments are availiable. In 
the Haldane-Wu analysis however, one starts from the idea that exotic statistics 
appears only as an interpolating statistics between bosons and fermions so that 
quantum states obeying generalized statistics play a secondary role only. We will 
not discuss here these two ways of viewing things; what we will do rather is to look 
for common features in both analysis. For example one ould like to know whether there 
exists a link between eqs.(1) and (3). What is the relation between the statistical
 weights in both approches and how exclusion statistics may be stated. In this paper we would like also to answer the open question of [19]concerning the link between
 the quasideterminant $\Delta _q$ used there and the quantum determinant $det_q$ 
 of non commutative geometry.\\
  
 The presentation of this paper is as follows: In 
 section 2, we study the generalized statistical wieght density and discuss the link 
 between eqs.(1) and (3) and the relation between the Haldane interpolating parameter 
 $\alpha$ and the spin. In section 3, we review the derivation of the wave function 
 of the gas of $N$ identical particles of spin $(1/M)$ $mod 1$ by using non commutative 
 geometric methods. Actually this section answers the open question arised in [19]. In section 4 we study the
  generalized Pauli exclusion principle for particles of spin $(1/M)$ $mod 1$;
  $M\geq 2$ by using q-antisymmetry. In section 5, we give our conclusion.

\section{The generalized statistical weight}

 First of all note that there are different ways of introducing generalized statistics. 
For the gas of $N$ particles with exclusion statistics we are interested in here, 
there are at least three ways of doing. The first way we will review brefly in this 
section, and which is due to Haldane, is based on the folowing combinatorial formula 
$W(\alpha)$ giving the number of accessible states of the gas of $N$ particles occupying 
a group of $G$ states
\be
 W(\alpha) = \frac{[G+(\alpha -1)(N-1)]!}{N![G- \alpha (N-1)]!},
\ee
In this equation, $\alpha$ is an interpolating parameter $0\leq \alpha \leq 1$ parametrising 
the generalized statistics. The second way of doing is that considered in [16,17,18].
It deals with quantum states with spin $(1/M)$ $mod 1$ and it is based on the conjencture that no more than
$(M-1)$ particles can live altogether on the same quantum state. The third way of doing was considered 
recently in [24]; it starts from postulating a formula for the wave function of the gas interpolating 
between the wave function of bosons and fermions by following the reasoning used for the densities eq.(4). 
For spin $(1/M)$ $mod 1$, the obtained wave function coincide with that given in [19]. Let us  now turn to 
explor the commun denominator between the Haldane and Wu approache and ours. First observe that eq.(4) may 
be rewritten as
\be
W(\alpha )=\frac {(x+y)!}{x!y!}=\frac {\Gamma (x+y+2)!}{\Gamma (x+1)\Gamma (y+1)},
\ee
where $x=N$; $y=y(\alpha)=G-\alpha N-(1-\alpha)$ and $\beta (x,y)$ is the well known digamma special function. 
Note that  $y(\alpha)$ is just the linear interpolation between the pionts $y_B=y(0)=G-1$ and $y_F=y(1)=G-N$ 
associated with the spin $s=0$ $mod 1$ and $s=1/2$ $mod 1$ respectively. this features as well as properties 
of the special function $\beta (x,y)$, in particular its integral definition, shows that the Haldane interpolating 
parameter $\alpha$ and the spin $s$ are related in the large $N$ limit as
\be
\alpha=\frac{N(1-2s)}{N-2}.
\ee
For the special case of fractional spins we are considering in this paper that is for spin $(1/M)$ $mod 1$, eq.(6) 
reduces to
\be
\alpha=\frac{N(M-2)}{M(N-2)}.
\ee
Taking $M=2$ and $M=N \to \infty$, we discover respectively the values $\alpha =0$ describing fermions and 
$\alpha =1$ for bosons. Moreover putting back eq.(7) into eq.(4), one gets the generalized statistical weight 
for spin $(1/M)$ $mod 1$. Actually this identity constitus the first bridge between our way of doing and the Haldane 
analysis. The second bridge we want to give concerns the link between eqs.(1) ans (3). Both of these distributions 
may be obtained by computing the partition function $\mathcal{Z}$ of the gas. Using the generalized Pauli exclusion 
principle according to which no more than $(M-1)$ particle can live altogether on the same quantum state. It is 
not difficult to check that the grand partition function $\mathcal{Z}$ is given by
\be
\begin{array}{lc}
\mathcal{Z}=\prod _{j\geq 0}\mathcal{Z_{j}}\\
\mathcal{Z_{j}}= \frac{1-exp-\beta M(\epsilon _j-M)}{1-exp-\beta (\epsilon _j-M)}.
\end{array} 
\ee
Eq.(8) extends the usual Bose and Fermi gas partition functions and leads to the mean occupation number of eq.(3). 
In the Haldane and Wu analysis, one uses the following property of the single level partition function $\mathcal{Z}(\xi)$
\be
\xi (\mathcal{Z}^\alpha (\xi)-\mathcal{Z}^{(\alpha -1)} (\xi))=1
\ee 
in order to calculate the mean number $\bar{n}=\xi \frac{\partial}{\partial \xi} log\mathcal{Z}(\xi) $. In doing 
so, one discovers that the mean number $\bar{n}$ obeys exactly eqs.(1-2). It should be noted that the explicit 
form of the quantum distribution (1) depends on the solving of eqs.(2). In [15], some special solutions were 
obtained. These conserns the Bose and Fermi distributions but also what is called there semions associated with 
$\alpha =1/2$ and having as mean number
\be
\bar{n}_j(\alpha =1/2)=\frac{1}{[\frac{1}{4}+exp2\beta (\epsilon _j -\mu)]^{1/2}}.
\ee
For general $\alpha$'s, the solution of eq.(2) is still missing. However if one requires that at most $(M-1)$ 
particles can be put per site with probabilities $1$, one descovers eq.(3).

\section{The generalized wave function}
In [17,19], an interpolating formula for the $\psi _{\{k_1,k_2,...,k_N \}}(1,2,...,N)$ wave function of a system 
of $N$ identical spin $(1/M)$ $mod 1$ particles was proposed. It is shown that $\psi$ is a kind of quasideterminant 
interpolating between the usual determinant and the permanent of fermions and bosons respectively. One of the 
remarkable features of the quasideterminat as defined in [18], see also [17], is the generalized Pauli exclusion 
principle manifests itself throught the identitiy
\be
[M]!=(1+q)(1+q+q^2)(1+q+q^2+....+q^{M-1})[1][2]....[M],
\ee
which vaniches identicaly for the roots $q=exp(\frac{2i\pi }{L})$ with $L=2,3,...,M$. For $M=2$, one recovers 
the Pauli exclusion principle and for $M=\infty$ we gets the Bose condensation. For generic values of $M$, 
$2\leq M\leq \infty $, one has just what we have been calling the generalized Pauli exclusion principle. The 
obtention of eq.(11) was in fact expexted since, on one hand it includes the q-fermionic number $Tr\,\, q^F$ 
which reads for spin $(1/M)$ $mod 1$ particles states as follows
\be
Tr\,\, q^F=1+q+q^2+.....+q^{M-1}
\ee
where $q^F=exp(\frac{2i\pi }{M})$. this equation extends the usual fermionic number and was shown in [8] 
to characterise the generalized Pauli exclusion principle. On the other hand, the introduction of the quasideterminant 
was mativated by earlier works on deformation of the conformal field theory in particular the $SU_k(2)/U(1)$ 
WSW model where we have the following identity
\be
(z_1-z_2)^{h_1+h_2-h}\phi _m^{j_1}(z_1)\phi _m^{j_2}(z_2)=(z_2-z_1)^{h_1+h_2-h}\phi _m^{j_2}(z_2)\phi _m^{j_1}(z_1).
\ee  
Eqs.(13) gives the short distance of the $2d$ fields operators $\phi _m^{j_1}(z_1)$ and $\phi _m^{j_2}(z_2)$
of conformal weights $h_1$ and $h_2$ respectively; $j$ is the $SU_k(2)$ isospin taking values in the range 
$0\leq j \leq \frac{k}{2}$; $-j\leq m\leq j$ is the U(1) Cartan charge and $h$ is a given highest weight which can 
be read from the fusion algebra of primary fields. For more details see [22]. Dividing both sides of eq.(12) 
by $(z_1-z_2)^{h_1+h_2-h}$, one gets a q-antisymmetric identity of deformation parameter $q=exp \pm i \pi(h_1+h_2-h)$. 
One the other hand, it was suggeted in [19] that eq.(11) and more generally the quasidetermiant $\Delta _q$ should 
be derived in a rigourous way by using non commutative geometric methods. However in trying to establish this connection, 
we were faced in the above mentionned works to a major difficulty which we had not solved at that time. The problem 
is that the quasideterminant $\Delta _q$ uses only commutative $\mathcal{C}$-number exactely as the wave function 
$\psi _{k_j} (j)$ of the individual particles whereas the quantum determinant $det_q$ is mainly based on a non 
commutative algebra. In this section we want to complete this result by establishing the link between $\Delta _q$ 
of [19] and the quantum determinant of non commutative geometry. To start let us recall brefly that the wave 
function $\psi$ of the gas (enclosed within a container volume $V$) is given by 
\be
\psi=\Delta _q (\psi _{k_1} (1), \psi _{k_2} (2), ....., \psi _{k_N} (N),
\ee
which for later use, we rewrite it as
\be
\psi =\Delta _q (\psi _j^N).
\ee
In this equation $\psi _{k_j} (j)$ stands for the wave fuction of the individual particle $j$. $\xi_j$ denote 
collectively all the coordinates of the $j-th$ particle namely its two position coordinates $(\tau , \sigma )$ 
and its spin $s$. $k_j$ is an index labeling its possible quantum states, in particular the energy $\epsilon _{k_j}$ 
the mementum $P_{k_j}$ of the particle and the orientation of the spin. Finally the quasideterminant is roughly 
speaking given by 
\be
\Delta _q(\psi _{k_j} (j))=\sum_{n=0}^{\frac{N(N-1)}{2}} q^n \,\, P_n \,(\psi _{k_j} (j)),
\ee
where $P_n \,(\psi _{k_j} (j))$ are homogenous polynomials in the individual wave functions. Details are given 
in [19]. Next, we introduce the $q$-deformed Grassman algebra generated by the system $\{\theta _1,\theta _2,....,\theta _N\}$ satisfying
\be
\begin{array}{lcl}
\theta _i^2 & = & 0\\
\theta _i \theta _j & = & q\, \theta_j \theta_i,\,\,\,\,i<j\\
\theta _i \theta _j & = & \frac{1}{q}\, \theta_j \theta_i,\,\,\,\,i>j,
\end{array} 
\ee
and consider the space of generalized wave function $\psi (j)$ defined as
\be
\psi (j)=\sum _{i=1}^N\,\,\psi _{k_j} (j)\,\theta _i.
\ee
In eq.(17-a), $\psi _{k_j} (j)$ are just the individual wave functions appearing in eq.(14). They may be viewed 
as $\mathcal{C}$-valued sections of a kind of a deformed $N$-dimentional one forms. For convenience, we prefer 
to rewrite eq.(17-a) as
\be
\psi (j)=\sum _{i=1}^N\,\,\psi _{ij} (j)\,\theta _i=[\psi][\theta],
\ee
where we have replaced $k_i$ by just the index $i$. Now, using eq.(16), one can work out the quantum determinant 
of $[\psi]$ by computing the generalized volume $V$ given by the product of all the superwave functions 
$\psi (j)$, $j=1,...,N$
\be
V=\prod _{j=1}^N\,\, \psi (j)\equiv det_q[\psi]\,.\,\prod _{j=1}^N\,\,\theta _j.
\ee          
Straightforward calculations show that $V$ reads as
\be
V=\Delta _q[\psi _{k_j} (j)]\prod _{i=1}^N\,\,\theta _i,
\ee
where $\Delta _q[\psi _{k_j} (j)]$ is exactly the quasideterminant we have introduced in [18] and wich we can define 
also as
\be
\Delta _q[\psi]=\sum_ {p \in S_N}q^{i(p)}\psi_{p(1)}(1)\psi_{p(2)}(2)....\psi_{p(N)}(N) .
\ee
In eq.(22) the summation is carried over all the permutations $p$ of the particles and $i(p)$ is the number 
of inversions of a given $p$. $i(p)$ is just the minimal lenght of $p$. For details sess [24], some useful features 
of eq.(20) will be considered in what follows.
\section{Generalized Pauli exclusion principle}

We begin by recalling that the Pauli exclusion principle manifests itself in different but equivalent ways. 
In the language of wave functions, the Pauli exclusion principle for fermions is carried by antisymmetry ensuring 
the vanishing of the Slater determinant whenever two fermions are in the same quantum state. In this section 
we want to show that this feature extends to particles with spin $(1/M)$ $mod 1$ satisfying a generalized Pauli 
exclusion principle. We will show in particular that this exclusion statistics is carried by a $q$ antisymmetry 
ensuring the vanishing of the wave function $\psi _{\{k_1,k_2,...,k_N \}}(1,2,...,N)$ whenever $M$ particles 
are in the same quantum state. To that purpose, let us first introduce the following convention notation
\be
\begin{array}{lcl}
\prod _{j=1}^N \psi(j) \equiv  det_q[\psi ^N].\prod _{n=1}^N \theta_n\\
\prod _{j=1}^N \psi(j) \equiv det_q[\psi _1 ^N].\prod _{n=1}^N \theta_n\\
\prod _{j=1}^N \psi(j) \equiv det_q[\psi _i ^N].\prod _{n=1}^N \theta_n,
\end{array} 
\ee 
In eq.(23) $[\psi _i^{(N-1)}]$, $j\ne i$ is just the $i$-th $(N-1 \times N-1)_{(i,i)}$ minor of $[\psi]$. Next 
using eqs.(16-17), it is not difficult to check that the above quantum determinant $det[\psi^N]$ and those of 
its minors $det(\psi^{(N-1)})$ are related as
\be
det_q(\psi_i^{(N-1)})=\sum_{j=1}^N \,\, q^{j-1} \psi _{k_N} (j) \,\, det_q(\psi _j^{(N-1)}),
\ee
or equivalently by expending along the $n$-th row
\be
det_q(\psi_i^{(N)})=\sum_{j=1}^N \,\, q^{N-j} \psi _{k_N} (j) \,\, det_q(\psi _j^{(N-1)}).
\ee
Note that eqs.(22) are just the generalisation of the Sarrus decomposition of the determinant. These equations 
which were conjenctured in [18,19] are as the basis of the derivation of the wave function for identical particles 
of spin $(1/M)$ $mod 1$. Note also that in the case where the $\psi _{k_i} (j)=\psi _{k_i}$ for all values of 
$j=1,....,N$; eq.(22-a) factorizes as
\be
det_q(\psi)=[N]!\psi _{k_1}\psi _{k_2}.....\psi _{k_N},
\ee
where $[N]!$ is as in eq.(11). Actually eq.(26) was first derived in [18] by using a different methods. 
It vanishes not only for $M=2$ $(q=1)$ as required by the Pauli exclusion principle but also for all values of 
$M$ lying between two and $N$. This means that the wave function $\psi _{\{k_1,k_2,...,k_N \}}(1,2,...,N)$ 
vanishes identically whenever $M$ collons of the quantum determinant $det_q(\psi^N)$ are equal. In order words, 
this eq.(26) means that no more than $(M-1)$ identical particle of spin $(1/M)$ $mod 1$ can live altogether on the 
same quantum state. This states then the generalized Pauli exclusion principle conjenctured in [16].

\section{conclusion}

In this paper we have studied features of generalized exclusion statistics by using different methods. We have 
also given the links between technics of statistical mechanics dealing with exotic statistics and our way of 
doing inspired from $2d$ conformal field theoretical methods. After a bref description on the ways one can follows 
to deal with a gas of identical particles with exotic spins, we show that the properties of the Haldane 
combinatorial formula eq.(4) are exactly those of the well known digamma $\beta (x,y)$
\be
\beta (x,y)=\int_0^1 \,\, t^{(x-1)} \,\, (1-t)^{(y-1)} \,\, dt.
\ee
Using standard features of $\beta (x,y)$, in particular the above integral definition, we have derived the link 
between the Haldane interpolating parameter $\alpha$ eq.(4) and the exotic spin $s$ of the particles of the gas  
as shown on eqs.(6-7). We have also established the link between the quantum distributions eqs.(1) and (3). We 
have shown that our formula (3) is also a solution of the Wu system (1) corresponding to the limiting case where 
at most $(M-1)$ particles can be put per state with probabilities $1$. We have also answered an open question 
rised in [18,19] asking for the link between our quasideterminant giving the wave function of the gas and the 
quantum determinant of non commutative geomerty. We have derived this link rigourously from $q$-anticommutativity. 
As a matter of facts we have derived some remarkable quantities which, we belive, has much to do with the generalized 
exclusion statistics. For spin $(1/M)$ $mod 1$, we can say that we have now a formula carrying the effect of 
the generalized Pauli exclusion principle namely
\be
[N]!=(1+q)(1+q+q^2)+......+(1+q+....+q^{M-1})+...+(1+q+....+q^{M-1}).
\ee
This equation includes an other quantity which was suggeted in [8] to describe this generalized principle. This 
concerns the $q$-fermionic number operator $Tr(q^F)$ eq.(12).\\

	\vskip 3cm
	{\bf Acknowledgment}\\
	One of the authors; J.Z. would like to thank the AS-ICTP and the High Energy Section 
	for hospitality. He would also to acknoledge the Student programm at ICTP for invitation.\\
	
	This research work has been supported by the programm PARS number: 372-98 CNR. 
        \newpage
       \vskip 1cm
{\bf References}
\begin{enumerate}
\item[[1]] B.I.Halperin, Phys. Rev. Lett. 52 (1984) 1583,\\
J.Frohlich, U.S.Studer and E.Thiram, in Proc. Conf. On three levels Lauven, (1993), eds M.Anes et al.
\item[[2]] I.Benkaddour and E.H.Saidi, Fractional supersymmetry as a matrix model, Class. Quant. Grav. 16(1999) 1793.
\item[[3]] J.M.Leinaas and J.Myrheim, Nuovo Cimenta B37 (1977) 1.
\item[[4]] A.Belavin, A.M.Polyakov and A.B.Zamolodchikov, Nucl. Phys. B241 (1984) 333.
\item[[5]] See for example, F.Wilczek, Fractional statistics and anyon superconductivity. (World scientific publishing) Singapou (1990) and refs therein.
\item[[6]] G.Musardo, Phys. Rep. 218 (1992) and refs therein. 
\item[[7]] L.D.Fadees, N.Yu.Reshetikhin and L.A. Takhtajan, Algebra analysis 1 (1987) 178.\\
M.Jimbo, T.Miwa and M.Odake, Lett. Math. Phys. 14 (1987) 123; Comm. Math. Phys. 116 (1988) 508.
\item[[8]] E.~H.Saidi, M.~B.~Sedra and J.~Zerouaoui, Class. Quant. Grav. 12 (1995) 1576.\\
    E.~H.Saidi, M.~B.~Sedra and J.~Zerouaoui, Class. Quant. Grav. 12 (1995) 2705.
\item[[9]] A.LeClair and C.Vafa, Nucl. Phys. B401 (1993) 413.
\item[[10]] J.Wess, Quantum groups and quantum spaces, LMU-TPW 97-18,\\
A.Connes, Non commutative geometry, Academic Press (1994).
\item[[11]] A.Connes, M.R.Douglas and A.Shwartz, JHEP 02 (1998) 03 hep-th/9711162.
\item[[12]] T.Banks, W.Fisher, S.H.Shenker and L.Susskind, Phys. Rev. D55 (1997) 5112, hep-th/96110043.
\item[[13]] S.Gokov, M.Rangamani and E.Witten, Dibaryon, string and brabes in ADS orbifold models, ASSNS-hep/98-93 ITEP-TH-30/98.
\item[[14]] F.D.H.Haldane, Phys. Rev. Lett. 67 (1991) 37.
 
\item[[15]] Y.S.Wu, Phys. Rev. Lett. 73, N7 (1994).
\item[[16]] E.H.Saidi, Procceding de la journee de Physique Statistique; Avril (1995), FST Settat, Morocco.
\item[[17]] . H.Saidi, ICTP preprint IC/95/299.\\
E.H.Saidi and J.Zerouaoui, Revue Marocaine des Sciences Physique N2 (2000).
\item[[18]] J.Zerouaoui, PHD thesis, Rabat Morocco (1995) and Doctorat d'Etat, Rabat Morocco (2000).
\item[[19]] M. Rachidi, E. H. Saidi and J. Zerouaoui, Phys. Lett. B409 (1997) 394.
\item[[20]] H.Chakir, A.Elfallah and E.H.Saidi, Mod. Phys. Lett. A (1995) 2931 and Class. Quant. Grav. 14 (1997) 2049.
\item[[21]] A.B.Zamolodchikov and V.A.Fateev, Sov. Phys. JETP 62 (1985) 215.
\item[[22]] P.C.Argyres, K.R.Dienes and S.H.H.Tye, Comm. Math. Phys. 154 (1993) 471.
\item[[23]] M.B.Sedra, PHD Thesis, Rabat Morocco (1993) and Doctorat d'Etat, Rabat Morocco (1995).
\item[[24]] M. Rachidi, E. H. Saidi and J. Zerouaoui, submited to JMP. 
\end{enumerate}
\end{document}